\documentclass[12pt,a4paper]{article}
\begin{document}
  
\title{Theory of acoustic analog of magneto-optic polar Kerr effect  under magnon-phonon resonance}

\author{A. M. Burkhanov, K. B. Vlasov, V. V. Gudkov, and B. V. Tarasov}

\maketitle

\section{Introduction }

\label{intr}

 The acoustic analog of magneto-optic Kerr effect consists of  variation of the polarization of the elastic wave after its reflection  from an interface between magnetic medium and isotropic non-magnetic one. Quantitative characteristics of the effects are  polarization parameters: $\varepsilon$ -- the ellipticity which modulus is the ratio of the minor and major ellipse axes and $\phi$ -- the angle of rotation of the polarization plane  or, more correctly,  of the major ellipse axis if  $\varepsilon\neq0$.

There are three possible variants  according to directions of the static magnetic induction {\bf B}, plane of incidence of the wave and the unit vector {\bf q} normal to the interface plane: polar (${\mathbf B}\parallel {\mathbf q}$); meridional (or longitudinal) ( ${\mathbf q}\times {\mathbf k} \perp {\mathbf B}$); and equatorial (or transverse) (${\mathbf q}\times {\mathbf k} \parallel {\mathbf B}$).  The  incident waves are called $p$-type if the elastic displacement vector ${\mathbf u}$ is perpendicular to ${\mathbf q}\times {\mathbf k}$, and $s$-type if ${\mathbf u}$ is parallel to this vector.

The acoustic analog of the Kerr effect was predicted by Vlasov and Kuleev \cite{vl_ku-68}. Theory of the  effect for an inclined incidence of a  wave was developed by Vlasov and Babushkin \cite{vl-bab_74}. In this paper an   isotropic magnetic medium (at ${\mathbf B}=0$) was discussed.

Here we consider the following model. The  medium I is a semi-infinite isotropic non-magnetic and non-dissipative while the medium II is  a ferromagnetic  one with cubic lattice and easy axis along  [111]-type crystallographic direction. The interface is perpendicular to the [001] axis and the plane of incidence is  given by the equation $x=0$. The  medium II has a  form of a sphere with  a small (respectively the radius of the sphere) plane area contacting with the medium I.

In this case the  elastic displacements of the reflected shear waves ${\bf u}^\prime$ may be represented by a sum of mutually orthogonal vectors ${\bf u}^\prime_{x}$ and ${\bf u}^\prime_{\perp}$, where
\begin{eqnarray}             
{\bf u}^\prime_{x}=u^\prime_{x}{\bf e}_x \:\:\:\:\:\:\:\:{\rm and}\:\:\:\:\:\:\:\:{\bf u}^\prime_{\perp}= u^\prime_{\perp} {\bf e}_\perp, \label{3bth26}
\end{eqnarray}
and  ${\bf e}_x$ and ${\bf e}_\perp$ are unit vectors. By introducing reflection coefficients $R^\pm_n$  for circular components defined by 
$R^\pm_n \equiv u^\prime_x/u_n \pm i u^\prime_\perp/u_n$
and expressing them in the form ${ R}^\pm_n=\left | { R}^\pm_n \right |\exp(i\rho\,^\pm_n)$, we obtain expressions for the ellipticity and the rotation of the polarization of an elastic wave of $n$-type upon reflection from the interface  as
\begin{equation}            
\varepsilon = \frac{\left | { R}^+_n \right | - \left | { R}^-_n \right |}{\left |{ R}^+_n \right | + \left |{ R}^-_n \right |}
\:\:\:\:\:\:\:\:{\rm and}\:\:\:\:\:\:\:\:\phi=\frac{1}{2}\left(\rho\,^-_n-\rho\,^+_n  \right). \label{3bth28}
\end{equation}
In definition (\ref{3bth26})  and later on a symbol with a prime  relates to characteristics of a reflected wave, with two primes -- to refracted (transmitted to the second medium) and without primes -- to the incident wave;   $n$ may refer to shear mode of arbitrary linear polarization.

\section{Boundary conditions}
\label{b-c}

 We will consider  classical approach  given in Ref.\ 3   and  generalize it to the case of magnetoelastic interaction in one of the contacting media. Thus, the boundary conditions require continuity of the displacements at the interface, and of the force acting on the interface:
\begin{equation}             
{\bf u}_n+\sum_{i=1}^3{\bf u}_i^\prime -\sum_{j=1}^m{\bf u}_j^{\prime\prime} =0, \label{pr4}
\end{equation}
\begin{equation}             
(\tau_{mz})_n+\sum_{i=1}^3(\tau_{mz})_i^{\prime} -\sum_{j=1}^m(\tau_{mz})_j^{\prime \prime} =0, \label{pr5}
\end{equation}
The sums in Eqs.\ (\ref{pr4})--(\ref{pr5}) contain contributions of all the normal (eigen) modes   that have elastic displacements. In harmonic approximation the  boundary conditions  can be presented as ones  for complex amplitudes and phases. The last one leads to the following conclusions: frequences of all the waves (incident, reflected, and refracted) are equal, 
 all the wave vectors  belong to a common plane, namely, the plane of incidence,  and, besides,
\begin{equation}             
\frac{\sin \theta}{s}=\frac{\sin \theta^{\prime}_i}{s^{\prime}_i}=\frac{\sin \theta^{\prime\prime}_j}{s^{\prime\prime}_j}, \label{3bth11}
\end{equation}
where $\theta$ is the angle of incidence, $\theta^{\prime}_i$ that of reflection, corresponding to the wave with phase velocity $s^\prime_i$, and $\theta^{\prime\prime}_j$ that of refraction, corresponding to the wave with phase velocity $s^{\prime\prime}_j$. 

 The boundary conditions given by Eqs.\ (\ref{pr4})--(\ref{pr5}) should be appended with ones for  variable magnetization those   were formulated in Ref.\ 1.

Remind:  all the eigen modes that have elastic displacements  should be represented in boundary conditions  and  the tensions due to the magnetoelastic interaction should be in Eq.\ (\ref{pr5}). 

 There are three  such solutions for an isotropic dielectric medium: one longitudinal and two degenerate transverse.

 For the magnetic medium, in general,  there should be five  solutions. However, here we will consider a  weak-coupling approximation.  In this approximation  the interaction between the subsystems results in small variation of the dispersion curves of normal modes which become coupled phonon-magnon (or phonon-like) and magnon-phonon (or magnon-like)  ones). Only three elastic-like modes can be accounted in Eqs.\ (\ref{pr4})--(\ref{pr5}) and  boundary conditions for magnetization are not used for obtaining $R_n^\pm$. 

\section{System of  equations for determination of normal modes}
\label{sy e}

To obtain the wave vectors  and complex amplitudes of the eigen modes  it is necessary to solve the equations of  the elasticity theory for the first medium and for the second one -- the system  consiting of Maxwell's equations, equation of  motion for magnetization, and equations  of elasticity theory. We propose that the crystal is in magnetically saturated state and only small variations of magnetization can occur.

Thus, for the first medium we have
\begin{equation}
\rho_1\ddot{u}_i=\frac{\partial \tau_{ij}}{\partial x_j}, \label{lj1a}
\end{equation}
with tensions $\tau_{ij}$ defined by
\begin{equation}   
\tau_{ij}=\frac{\partial  W}{\partial \varepsilon_{ij}}  , \label{lj2a}
\end{equation}
where $\rho_1$ is the density of the medium I, $\varepsilon_{ij}=\frac{1}{2} \left( {\partial u_i}/{\partial x_j}+ {\partial u_j}/{\partial x_i}\right)$ are components of strain (or deformation) tensor,  and the density of free energy $W$ contains only   the elastic     energy $W_L$. 

For the second medium we have equation of motion for volume element similar to  (\ref{lj1a}). However, the free energy which defines the  tensions  and effective field {\bf H}$_e$ should be presented by elastic, magnetic,  and magnetoelastic terms. 

Formally, we should use the superscript $\prime\prime$ on all of the parameters and variables related to the magnetic medium, but we will defer this until such terms for all of the waves, incident, reflected, and transmitted, appear in the same expression.

It is preferable to write the equation of motion for magnetization in the form of the Gilbert equation:
\begin{equation}   
\dot{\bf M}=\gamma {\bf M} \times {\bf H}_e+\frac{\alpha_0}{M_0}{\bf M} \times \dot{\bf M},  \label{3bth23a}
\end{equation}
where  $\alpha_0$ is the relaxation parameter, $M_0$ is the modulus of  static magnetization in homogeous  state,   and  ${\mathbf M}$ is total magnetization. 

The density of  total free energy of a magnetic crystal  in addition to elastic and magnetoelastic energy, contains  Zeeman energy $W_0$, exchange energy $W_{ex}$, energy of  magnetic anisotropy $W_a$,  and magnetostatic energy  $W_d$. The last two terms determines the field of magnetic anisotropy ${\mathbf H}_a$   and the static  ${\mathbf H}_d^0$   and variable ${\mathbf h}$ demagnetizing fields.  We will use demagnetizing factor  corresponding to a spherical specimen considering the plane areas as small. Expressions for the types of energy and fields  can be found in  Ref.\ 4.

Since we are limiting ourselves to the case where all wave vectors lie in the $y-z$ plane, the complete system has the form
\begin{eqnarray}
a_{11}u_1 + a_{14}m_1 = 0, \nonumber \\
a_{22}u_2 + a_{23}u_3 + a_{25}m_2 = 0, \nonumber \\
a_{32}u_2 + a_{33}u_3 + a_{35}m_2 = 0, \label{3bth31} \\
a_{42}u_2 + a_{43}u_3 + a_{44}m_1 + a_{45}m_2 = 0, \nonumber \\
a_{51}u_1 + a_{54}m_1 + a_{55}m_2 = 0, \nonumber
\end{eqnarray}
where ${ m_i}$ are components of  variable magnetization,
\begin{eqnarray}
a_{11} = \rho\omega^2-c_{1313}(k^2_2+k^2_3),\:\:
  a_{22} = \rho\omega^2-c_{1111}k^2_2-c_{1313}k^2_3,\nonumber \\
a_{33} = \rho\omega^2-c_{1111}k^2_3-c_{1313}k^2_2, \:\: a_{44}=a_{55}=i\omega, \nonumber\\
a_{14} = a_{25} = -\frac{i b_2 k_3}{M_0},\:\: 
 a_{23} = a_{32} = -(c_{1122} +c_{1313})k_2k_3,\nonumber\\
a_{35} = -\frac{i b_2 k_2}{M_0}, \:\: a_{42} = -a_{51} = i\gamma b_2 k_3, \nonumber\\
a_{43} = i\gamma b_2 k_2, \:\: a_{54} = \gamma\left[\frac{2A}{M_0}(k_3^2+k_2^2)+ H_i\right]-i \alpha_0 \omega, \nonumber\\
{a_{45} = -\gamma\left[\frac{2A}{M_0}(k_3^2+k_2^2)+ H_i +\frac{4\pi k_2^2 M_0}{k_2^2 + k_3^2}\right]+i \alpha_0 \omega}, \nonumber  
\end{eqnarray}
$\rho$ is the density of the medium II,   $H_i=H+H_d^0+H_a$, $c_{ijkl}$ are elastic moduli, $ b_2$  is magnetoelastic constant.  It was assumed that all variables are proportional to $\exp\left[ i \left(\omega t- {\bf k}\cdot {\bf r} \right)\right]$.

Existing of non-trivial solutions of the system (\ref{3bth31}) requires its determinant to be equal to the zero. It is  dispersion equation since it  gives expressions for the wave vectors of eigen modes. As the angle of incidence is given and therefore, according to Eq.\ (\ref{3bth11}),  Re($k_2$) for all of the eigenvectors is defined, the unknown values are  Re($k_3$), Im($k_3$),  and Im($k_2$).

\section{Solution of dispersion equation}
\label{A1}

Using the last two equations of (\ref{3bth31}), we may express $m_1$ and $m_2$ as functions of $u_1$, $u_2$, and $u_3$:
\begin{eqnarray}             
m_i=b_{ij}u_j,  \label{3bth34a}
\end{eqnarray}
 substitute these expressions into the first three equations, and set the determinant of the new system (containing these three equations) equal to zero:
\begin{eqnarray}
\lefteqn{
\left(a_{11}+a_{14}b_{11} \right)} \nonumber \\& & \times 
\left[\left( a_{22}+a_{25}b_{22}\right)\left(a_{33}+a_{35}b_{23} \right)-\left(a_{32}+a_{35}b_{22} \right)\left(a_{23}+a_{25}b_{23} \right)\right] \nonumber \\
& & \hspace{0cm} \mbox{} -a_{14}b_{12}\left[ a_{25} b_{21} \left( a_{33} + a_{35}b_{23} \right) - a_{35} b_{21} \left( a_{23} + a_{25}b_{23}\right) \right] \nonumber \\
& & \hspace{0cm} \mbox{} +a_{14} b_{13}\left[a_{25}b_{21}\left(a_{32}+a_{35} b_{22}  \right)- a_{35}b_{21} \left( a_{22} +a_{25}b_{22}\right)\right]=0, 
\label{3bth33}
\end{eqnarray}
where
\begin{eqnarray}
b_{11}  = \frac{a_{45}a_{51}}{a_{44}a_{55}-a_{54}a_{45}},\:\:  b_{12}  = \frac{-a_{55}a_{42}}{a_{44}a_{55}-a_{54}a_{45}},\nonumber \\  b_{13}  = \frac{-a_{55}a_{43}}{a_{44}a_{55}-a_{54}a_{45}}, \:\:
b_{21}  = \frac{-a_{51}a_{44}}{a_{44}a_{55}-a_{54}a_{45}},\label{3bth34} \\ 
b_{22}  = \frac{a_{54}a_{42}}{a_{44}a_{55}-a_{54}a_{45}},\:\:  b_{23}  = \frac{a_{43}a_{54}}{a_{44}a_{55}-a_{54}a_{45}}, \nonumber  
\end{eqnarray}

We will solve the dispersion equation  (\ref{3bth33}) by an iterative method. The zero order approximation will be obtained by letting the elements describing the magnon-phonon interaction (i.e., $a_{14}$, $a_{25}$, $a_{35}$, $a_{41}$, $a_{42}$, $a_{43}$, $a_{51}$, $a_{52}$, and $a_{53}$) vanish. One of the solutions is 
\begin{equation}
k_3^2={\rho \omega^2}{c_{1313}^{-1}}-k^2_2,  \label{3bth35}
\end{equation}
whereas the next two are
\begin{eqnarray} 
k_3^2=-{P}\pm\left({P^2}-Q\right)^{1/2}, \label{3bth36}
\end{eqnarray}
where
\begin{eqnarray}
P\, &\!\!\! = &\!\!\! \,\left[ \left(c_{1111}^2+c_{1313}^2-\left(c_{1122}+c_{1313}\right)^2\right)k_2^2\right.\nonumber\\ & & \mbox{\hspace{0.5cm}}\left.-\rho\omega^2\left(c_{1313}+c_{1111}\right)\right]\left(2c_{1111}c_{1313}\right)^{-1},   \nonumber\\
Q\, &\!\!\! = &\!\!\! \,\rho^2\omega^4 -\rho\omega^2k_2^2\left( c_{1313}+c_{1111}\right)+c_{1313}c_{1111}k_2^4, \nonumber
\end{eqnarray}

Eqs.\ (\ref{3bth35})--(\ref{3bth36}) for $k_3$ relate to pure elastic waves:  transverse $s$-type mode with wave vector ${\bf k}_s$,  quasi-transverse (quasi-$p$) mode with wave vector ${\bf k}_p$, and  quasi-longitudinal (quasi-$l$) mode with wave vector ${\bf k}_l$. At this point we have obtained real wave vectors (i.e., their imaginary parts are zero), since energy dissipation is considered to be in the magnetic subsystem only.

Next, the resulting expressions for $k_3$ should be substituted in turn into the equations which describe the magnetic subsystem and the magnon-phonon interaction. Recall that all $k_2$ in the zeroth approximation are real, equal, and known. Thus the solution for $k_3$ relating to the first weakly-coupled phonon-magnon mode  (now of the quasi-$s$ type, since it has $y$ and $z$ components of the elastic displacement) is
\begin{eqnarray}
(k_3)^2_{qs}=- k_2^2+\frac{1}{c_{1313}}\left[{\rho \omega^2}+a_{14}b_{11} +\left[\left( a_{22}+a_{25}b_{22}\right)\left(a_{33}+a_{35}b_{23} \right)
\right.\right.
\nonumber\\
\left.\left.-\left(a_{32}+a_{35}b_{22} \right)\left(a_{23}+a_{25}b_{23} \right)\right]^{-1} \right.\nonumber\\
\left.\times \left[ a_{14} b_{13}\left[a_{25}b_{21}\left(a_{32}+a_{35} b_{22}  \right)- a_{35}b_{21} \left( a_{22} +a_{25}b_{22}\right)\right]\right.\right.\nonumber\\ \left.\left. -a_{14}b_{12}\left[ a_{25} b_{21} \left( a_{33} + a_{35}b_{23} \right) - a_{35} b_{21} \left( a_{23} + a_{25}b_{23}\right) \right]\right]\right],  \label{3bth37b}
\end{eqnarray}
where all $a_{ij}$\ and $b_{mn}$\  are functions of $(k_3)_s$.

The other two solutions relating to the coupled phonon-like  modes are written in the form of (\ref{3bth36}), however the parameter $Q$ should now be given by
\begin{eqnarray}
Q= \rho^2\omega^4 -\rho\omega^2\left[ k_2^2\left( c_{1313}+c_{1111}\right) \right]
+c_{1313}c_{1111}k_2^4 +a_{35} a_{22}b_{23} +a_{25} a_{33}b_{22}\nonumber\\
-a_{32} a_{25}b_{23} -a_{35} a_{23}b_{22} -a_{14}\left[ a_{11}+ a_{14} b_{11} \right]^{-1}\nonumber\\
\times \left[ b_{12}\left[a_{25}b_{21}
\left(a_{33}+a_{35}b_{23} \right) - a_{35}b_{21}
\left(a_{23}+a_{25}b_{23} \right) \right]\right.\nonumber\\ \left.  - b_{13}\left[a_{25}b_{21} \left(a_{32}+a_{35}b_{22} \right) -
 a_{35}b_{21}
\left(a_{22}+a_{25}b_{22} \right) \right]\right]. \label{3bth38}
\end{eqnarray}
Note that the solution of Eq.\ (\ref{3bth36}) with the $+$ sign before the root describes the coupled quasi-$p$ mode for which all $a_{ij}$ and $b_{kn}$ in Eq.\ (\ref{3bth38}) are functions of $(k_3)_p$, whereas that with the $-$ sign describes the coupled quasi-$l$ mode for which all $a_{ij}$ and $b_{kn}$ are functions of $(k_3)_l$.

Having obtained the complex $z$ components of the wave vectors, we can now determine the imaginary part of the $y$ components for a particular ${\bf k}=k{\bf e}_{\bf k}$, where $k$ is a complex quantity and ${\bf e}_{\bf k}$ is a unit vector in the ${\bf k}$ direction. Thus, the definition ${\rm Re}(k_i)={\rm Re}[k\cos({\bf k}\cdot{\bf e}_i)]$ should also be correct for the imaginary parts of the wave vector components. For every ${\mathbf k}_{qj}$  it follows that
\begin{eqnarray}
{\rm Im}(k_2)_{qj} = {\rm Im}(k_3)_{qj}\frac{\cos({\mathbf k} \cdot {\mathbf e}_3)}{\cos({\mathbf k} \cdot
{\mathbf e}_2)}
 = {\rm Im}(k_3)_{qj}\frac{{\rm
Re}(k_2)_{qj}}{{\rm Re}(k_3)_{qj}},\nonumber
\end{eqnarray}
where the index $q$ stands for {\it quasi-}  and $j$  for $s, p,$ or $l$.

To illustrate the differences in the positions of magnon-phonon resonances for waves of different polarizations and to compare them with data for $\theta=0$, Fig.~1 shows the resonant frequency $\omega_r$ versus magnetic field
calculated for a YIG single crystal. This frequency is determined by requiring that the real part of the resonant denominator $a_{44}a_{55}-a_{54}a_{45}$ in the definitions of $b_{ij}$ given by  expresions (\ref{3bth34}) vanish. Recall that $\theta$ is the angle of incidence, while the angle of refraction  for $\theta \neq 0$ depends on $H$ at a particular frequency and therefore cannot be a fixed property of curves {\sl B--D}.  It can be seen that in general one can observe manifestation of three resonances in the properties of a reflected wave.

\begin{figure}
\begin{picture}(155,100) 
\put(15,0){\special{eps:figure1.eps x=11.15cm y=9.cm}}
\end{picture}
\caption{Magnetic field dependences of the resonant frequency for waves propagating in a YIG single crystal with {\bf H} parallel to [001]. Curve {\sl A} corresponds to a transverse wave propagating along $\bf H$ (normal incidence at the interface), curves {\sl B}, {\sl C}, and {\sl D} -- to quasi-$s$, quasi-$p$, and quasi-$l$ modes, respectively, calculated for the angle of incidence $\theta=22.5^\circ$.}

\label{resfre} 
\end{figure}

\section{Complex amplidutes of  vibrations and tensions}
\label{ca}

The complex amplitudes of the elastic vibrations and tensions (elastic and magnetoelastic)  can be written by inserting the eigen wave vectors into the system (\ref{3bth31}). In both cases magnetization is expressed in terms of $u_1$, $u_2$, and $u_3$ according to Eq.\ (\ref{3bth34a}) .

The first three equations of (\ref{3bth31}), written in the form
\begin{eqnarray}
c_{mn}u_n=0 \:\:\:(m,n=1,2,3)
\label{3bth38b} 
\end{eqnarray}
 yield complex amplitudes expressed in terms of those which do not vanish for $b_2\to 0$ and $\theta\to 0$ (i.e., in terms of $U_{1s}^{\prime\prime}$ for quasi-$s$ mode, $U_{2p}^{\prime\prime}$ for quasi-$p$, and $U_{3l}^{\prime\prime}$ for quasi-$l$; we  omit the   $q$ index here and later on for complex amplitudes and wave vectors in magnetic medium),   $c_{11}  = a_{11}+a_{14}b_{11},  c_{12}  = a_{14}b_{12},  c_{13}  = a_{14}b_{13}, 
c_{21}  = a_{25}b_{21}, c_{22}  = a_{22}+a_{25}b_{22},  c_{23}  = a_{23}+a_{25}b_{33}, 
c_{31}  = a_{35}b_{21}, c_{32}  = a_{32}+a_{35}b_{22},  c_{23}  = a_{33}+a_{35}b_{33}$, $b_{ij}$ are defined by (\ref{3bth34}).

So in the magnetic medium we have the following expressions for the complex amplitudes of  phonon-like quasi-$s$, -$p$, and -$l$ modes where we denote their wave vectors by ${\bf k}_s^{\prime\prime}$, ${\bf k}_p^{\prime\prime}$, and ${\bf k}_l^{\prime\prime}$:
\begin{eqnarray}
U_{2s}^{\prime\prime}  = \frac{c_{31}c_{23}-c_{21}c_{33}}{c_{22}c_{33}-c_{32} c_{23}}U_{1s}^{\prime\prime}\equiv f_{1}({\bf k}_s^{\prime\prime}) U_{1s}^{\prime\prime}\equiv A_{21}U_{1s}^{\prime\prime}, \nonumber\\
U_{3s}^{\prime\prime} = \frac{c_{32}c_{21}-c_{22}c_{31}}{c_{22}c_{33}-c_{32} c_{23}} U_{1s}^{\prime\prime}\equiv f_{2}({\bf k}_s^{\prime\prime}) U_{1s}^{\prime\prime} \equiv A_{31}U_{1s}^{\prime\prime},\nonumber\\
U_{1p}^{\prime\prime}  =\frac{c_{32}c_{13}-c_{12}c_{33}}{c_{11}c_{33}-c_{31} c_{13}}U_{2p}^{\prime\prime}\equiv f_{3}({\bf k}_p^{\prime\prime}) U_{2p}^{\prime\prime}\equiv A_{12}U_{2p}^{\prime\prime}, \nonumber\\
U_{3p}^{\prime\prime}  = \frac{c_{31}c_{12}-c_{11}c_{32}}{c_{11}c_{33}-c_{31} c_{13}} 
U_{2p}^{\prime\prime}\equiv f_{4}({\bf k}_p^{\prime\prime}) U_{2p}^{\prime\prime} \equiv A_{32}U_{2p}^{\prime\prime},
\label{3bth39c}\\U_{1l}^{\prime\prime}  = \frac{c_{23}c_{12}-c_{13}c_{22}}{c_{11}c_{22}-c_{21} c_{12}}U_{2p}^{\prime\prime}\equiv f_{5}({\bf k}_l^{\prime\prime}) U_{3l}^{\prime\prime}\equiv A_{13}U_{3l}^{\prime\prime}, \nonumber\\
U_{2l}^{\prime\prime}  = \frac{c_{21}c_{13}-c_{11}c_{23}}{c_{11}c_{22}-c_{21} c_{12}} U_{2p}^{\prime\prime}\equiv f_{6}({\bf k}_l^{\prime\prime}) U_{3l}^{\prime\prime} \equiv A_{23}U_{3l}^{\prime\prime}\nonumber.
\end{eqnarray}

In the non-magnetic medium we have an incident wave  of $p$-type and  reflected waves of $s$-, $p$-, and $l$-types with the angles of reflection   defined by  Eq.\ (\ref{3bth11}).

The complex amplitudes of the tensions  associated with quasi-$s$, quasi-$p$, and quasi-$l$ waves  can be written in the form
\begin{eqnarray}
(\tau_{13}^0)^{\prime\prime}_s = \left[-i c_{1313}(k_3^{\prime\prime})_s+ \frac{b_2}{M_0}\left(b_{11} + b_{12}A_{21} + b_{13}A_{31}\right)\right] U^{\prime\prime}_{1s}
\equiv (T_{13})_s^{\prime\prime}U^{\prime\prime}_{1s},\nonumber\\
(\tau_{23}^0)^{\prime\prime}_s = \left\{-ic_{2323}\left[(k_2^{\prime\prime})_sA_{31} + (k_3^{\prime\prime})_s A_{21} \right] \right.{\hspace{4cm}}\nonumber\\ \left.+ \frac{b_2}{M_0}\left( b_{21} + b_{22}A_{21}+ b_{23}A_{31}\right)\right\} U^{\prime\prime}_{1s}
 \equiv (T_{23})_s^{\prime\prime} U^{\prime\prime}_{1s}, \nonumber \\
(\tau_{33}^0)^{\prime\prime}_s = -i\left[c_{3322}(k_2^{\prime\prime})_sA_{21} +c_{3333}(k_3^{\prime\prime})_sA_{31} \right] U^{\prime\prime}_{1s}\equiv (T_{33})_s^{\prime\prime} U^{\prime\prime}_{1s}, \nonumber\\*[6pt]
(\tau_{13}^0)^{\prime\prime}_p = \left[-ic_{1313}(k_3^{\prime\prime})_pA_{21} + \frac{b_2}{M_0}\left( b_{11} A_{12}+ b_{12}+ b_{13}A_{32} \right)\right]
U^{\prime\prime}_{2p} 
\nonumber\\ 
 \equiv  (T_{13})_p^{\prime\prime} U^{\prime\prime}_{2p},\nonumber   \\*[6pt]
(\tau_{23}^0)^{\prime\prime}_p = \left\{ -ic_{2323}\left[(k_2^{\prime\prime})_pA_{32} +(k_3^{\prime\prime})_p \right]+ \frac{b_2}{M_0}\left( b_{21}A_{12}+b_{22}+b_{23}A_{32} \right) \right\}U^{\prime\prime}_{2p}\nonumber\\ 
 \equiv  (T_{23})_p^{\prime\prime} U^{\prime\prime}_{2p},\nonumber \\*[6pt]
(\tau_{33}^0)^{\prime\prime}_p = -i\left[c_{3322}(k_2^{\prime\prime})_p+c_{3333}(k_3^{\prime\prime})_p A_{32} \right] U^{\prime\prime}_{2p}\nonumber\\ 
\equiv (T_{33})_p^{\prime\prime} U^{\prime\prime}_{2p},\nonumber\\*[6pt]
(\tau_{13}^0)^{\prime\prime}_l = \left[ -ic_{1313}(k_3^{\prime\prime})_l A_{13}+ \frac{b_2}{M_0}\left(b_{11}A_{13} + b_{12}A_{23}+b_{13} \right)\right]U^{\prime\prime}_{3l}\nonumber\\ 
 \equiv  (T_{13})_l^{\prime\prime}U^{\prime\prime}_{3l}, \nonumber \\*[6pt]
(\tau_{23}^0)^{\prime\prime}_l = \left\{-i c_{2323}\left[(k_2^{\prime\prime})_l+ (k_3^{\prime\prime})_lA_{23} \right]+ \frac{b_2}{M_0}\left( b_{21}A_{13} +b_{22}A_{23}+b_{23} \right) \right\}U^{\prime\prime}_{3l} \nonumber\\ 
\equiv  (T_{23})_l^{\prime\prime}U^{\prime\prime}_{3l}, \nonumber\\*[6pt]
(\tau_{33}^0)^{\prime\prime}_l =\, -i\left[ c_{3322}(k_2^{\prime\prime})_l\, A_{23} + c_{3333} (k_3^{\prime\prime})_l \,\right]U^{\prime\prime}_{3l}\equiv (T_{33})_l^{\prime\prime}U^{\prime\prime}_{3l}. \nonumber\\\label{3bth46}
\end{eqnarray} 
Note that all of the coefficients $(T_{ab})_j^{\prime\prime}$ are functions of ${\bf k}_j^{\prime\prime}$, where $j$ indicates the type of the wave (quasi-$s$,\ -$p$,\ or\ -$l$) in the magnetic medium.

The tensions in the non-magnetic medium may be written in terms  of the Lam\'e constants $\lambda$ and $\mu$ . 

\section{Ellipticity and rotation of the polarization}
\label{r-c}

To obtain the amplitudes of the reflected waves, we have to write Eqs.\ (\ref{pr4})--(\ref{pr5})    in the form:
\begin{eqnarray}    
\frac{U^{\prime\prime}_{1s}}{U_p}+ A_{12}\frac{U^{\prime\prime}_{2p}}{U_p}+ A_{13}  \frac{U^{\prime\prime}_{3l}}{U_p}  -\frac{U^\prime_p}{U_p} =0\nonumber\\
A_{21}\frac{U^{\prime\prime}_{1s}}{U_p} +\frac{U^{\prime\prime}_{2p}}{U_p} + A_{23} \frac{U^{\prime\prime}_{3l}}{U_p} -\cos\theta\, \frac{U^\prime_p}{U_p} +\sin\theta^\prime_l \,U^\prime_l/U_p=\cos\theta \nonumber\\
A_{31}\frac{U^{\prime\prime}_{1s}}{U_p} + A_{32} \frac{U^{\prime\prime}_{2p}}{U_p}+  \frac{U^{\prime\prime}_{3l}}{U_p}  -\sin\theta \,\frac{U^\prime_p}{U_p}  -\cos\theta^\prime_l \,\frac{U^\prime_l}{U_p }=   - \sin\theta \nonumber \\
(T_{13})_s^{\prime\prime} \frac{U^{\prime\prime}_{1s}}{U_p} + (T_{13})_p^{\prime\prime} \frac{U^{\prime\prime}_{2p}}{U_p} + (T_{13})_l^{\prime\prime} \frac{U^{\prime\prime}_{3l}}{U_p}  - i\mu \frac{\omega}{s_p}\cos\theta \frac{U^\prime_s}{U_p }=   0  \label{sys}\\
(T_{23})_s^{\prime\prime} \frac{U^{\prime\prime}_{1s}}{U_p} + (T_{23})_p^{\prime\prime} \frac{U^{\prime\prime}_{2p}}{U_p} + (T_{23})_l^{\prime\prime}  \frac{U^{\prime\prime}_{3l}}{U_p}{\hspace{3cm}}\nonumber\\  -i \mu\frac{\omega}{s_p}\cos 2\theta   \frac{U^\prime_p}{U_p}  +i \mu\frac{\omega}{s_l}\sin 2\theta^\prime_l \frac{U^\prime_l}{U_p } = -i \mu\frac{\omega}{s_p}\cos 2\theta \nonumber\\
(T_{33})_s^{\prime\prime} \frac{U^{\prime\prime}_{1s}}{U_p} + (T_{33})_p^{\prime\prime} \frac{U^{\prime\prime}_{2p}}{U_p} + (T_{33})_l^{\prime\prime}  \frac{U^{\prime\prime}_{3l}}{U_p}  - 2\, i \mu \frac{\omega}{s_p}\sin 2\theta   \frac{U^\prime_p}{U_p}{\hspace{5mm}}\nonumber\\ - \left[2\,i \frac{\omega}{s_l} \left(\lambda+2\mu \cos^2\theta_l^\prime\right)\right] U^\prime_l/U_p=2\, i \mu \frac{\omega}{s_p}\sin 2\theta \nonumber
\end{eqnarray}

Finally, we have a system of six inhomogeneous linear equations in six unknowns. Our particular interest is in $U^\prime_s/U_p$\ and $U^\prime_p /U_p$ (note $U^\prime_s$ and $U_p^\prime$ are complex amplitudes of $u^\prime_{x}$ and $u^\prime_{\perp}$, respectively, in the definition (\ref{3bth26})), which yield the reflection coefficients in terms of complex amplitudes $R^\pm_p =U^\prime_s/U_p\pm i\,U^\prime_p /U_p$. 

Thus, after solving the system (\ref{sys}), we have obtained all the data required to determine $\varepsilon$\ and $\phi$\ from Eqs.\  (\ref{3bth28}).

To illustrate the phenomena, Fig.~2 shows the Kerr rotation and ellipticity  which occur at a quartz-YIG interface.  Calculations performed on the basis of the scheme given in section~\ref{A1} show that variations of the wave vector related to a quasi-$s$ type mode exceed those of quasi-$p$ and quasi-$l$ modes by factors of $10^2$ and $10^3$, respectively. As a consequence, only the $z$ component of $\Delta {\bf k}_s^{\prime\prime}$ is shown in this figure to compare with the behaviors of $\varepsilon(H)$ and $\phi(H)$. Remember  we assumed that the ferromagnet is in a magnetically homogeneous state, thus the results are valid only above the field of magnetic saturation, $H_S=667$ Oe. In Fig.~2 (b), (c), (e), and (f) these parts of the curves are shown as solid lines.  It follows that only the high field wing of the curves can be observed at the frequency of $53.5$ MHz with the relaxation parameter $\alpha_0>0.1$.

\begin{figure}
\begin{picture}(155,180) 
\put(00,120){\special{eps:figure2a.eps x=7.5cm y=6.cm}}

\put(00,60){\special{eps:figure2b.eps x=7.5cm y=6.cm}}

\put(00,00){\special{eps:figure2c.eps x=7.5cm y=6.cm}}

\put(75,120){\special{eps:figure2d.eps x=7.5cm y=6.cm}}

\put(75,60){\special{eps:figure2e.eps x=7.5cm y=6.cm}}
-
\put(75,00){\special{eps:figure2f.eps x=7.5cm y=6.cm}}
\end{picture}
\caption{(a), (d) Magnetic field dependences of the real ({\sl A}) and imaginary ({\sl B}) parts of the $z$ component of the wave vector of a quasi-$s$ phonon-like mode  with a frequency of $53.5$ MHz propagating in a YIG single crystal for an angle of incidence   $\theta=22.5^\circ$  and  rotation of the polarization  (b), (e) and the ellipticity (c), (f) of the reflected wave calculated with $\alpha_0=0.5$ for (a)--(c) and $0.1$ for (d)--(f); $\Delta {\bf k}_s \equiv {\bf k}_s^{\prime\prime}(H)-{\bf k}_s^{\prime\prime}(0)$.  }
\label{YIG_Kerr_a} 
\end{figure}

\section* {Acknowledgement}

The authors  acknowledge  financial support from  International Science Foundation (grant RGD000--RGD300).

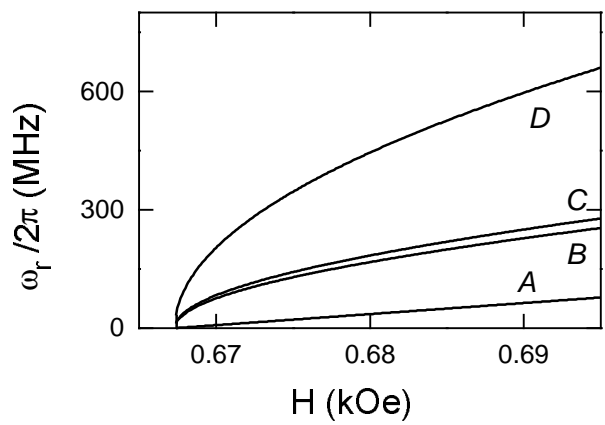
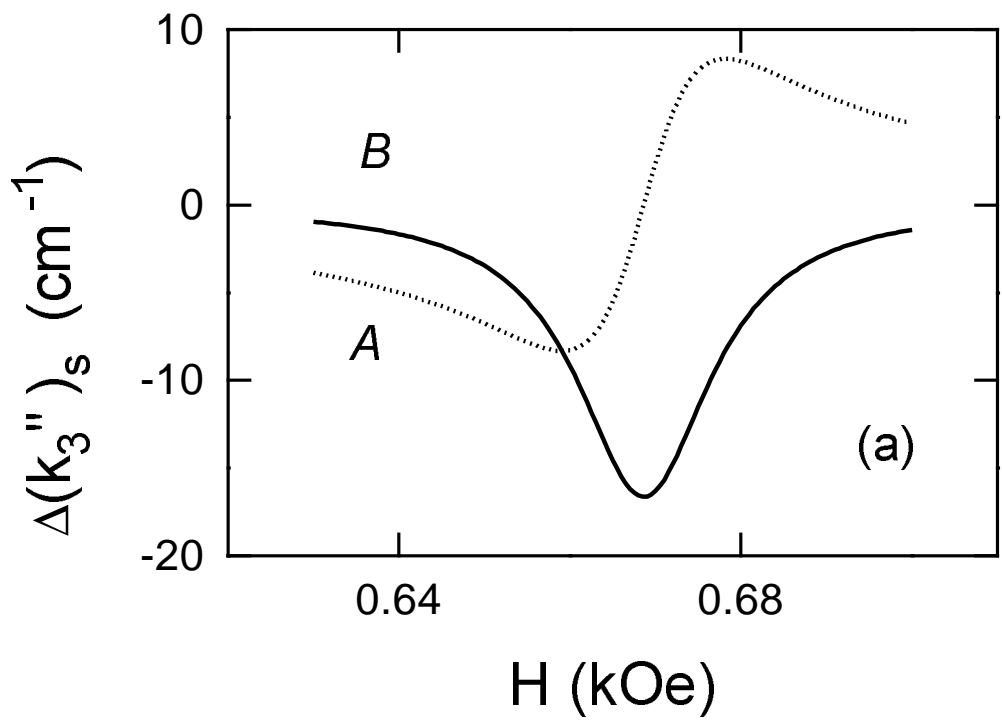
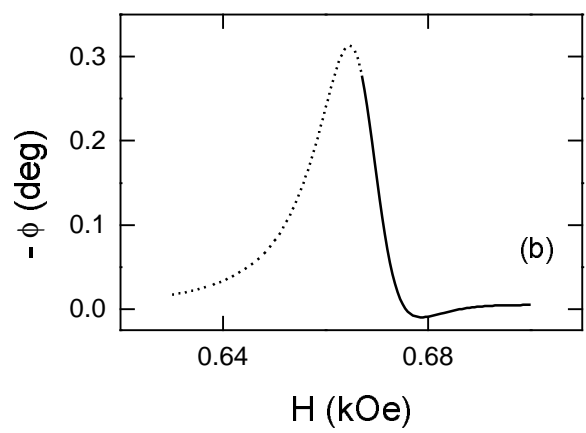
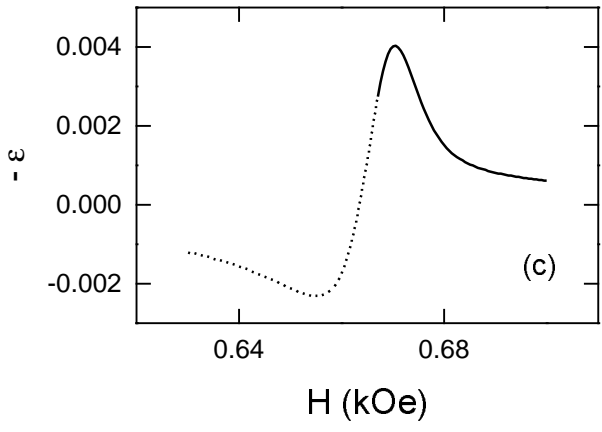
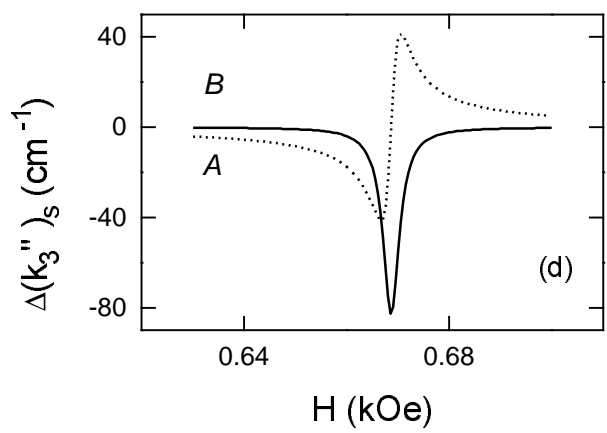
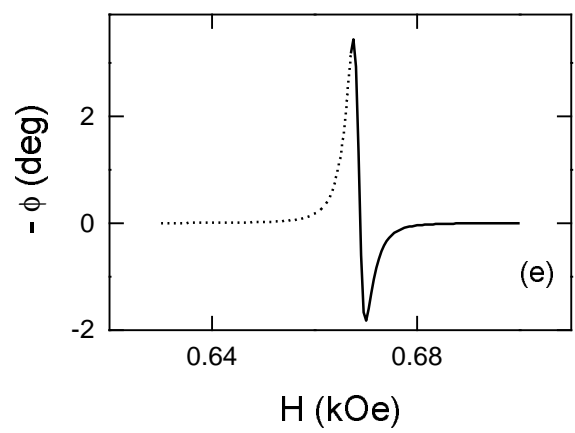
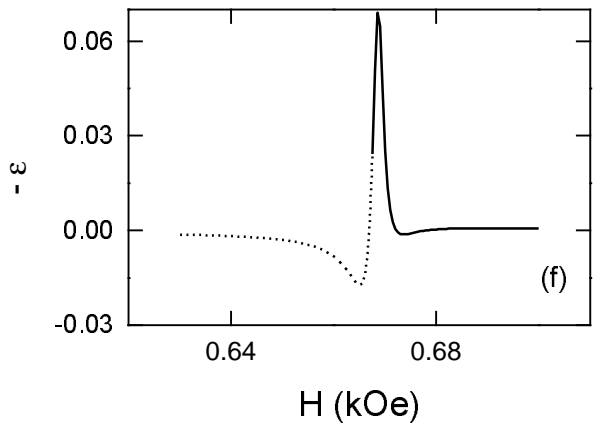
\end{document}